# Fractal universe and the speed of light: Revision of the universal constants

Antonio Alfonso-Faus

E.U.I.T. Aeronáutica
Plaza Cardenal Cisneros 40, 28040 Madrid, Spain
E-mail: aalfonsofaus@yahoo.es

**Abstract.** We apply the property of selfsimilarity that corresponds to the concept of a fractal universe, to the dimension of time. It follows that any interval of time, given by any tick of any clock, is proportional to the age of the universe. The fractality of time gives the fractality of space and mass. First consequence is that the speed of light decreases inversely proportional to time, same as the Hubble parameter. We then revise the universal constants and, at the cosmological scale, they are all of order one, as Dirac proposed. We find three different scales, each one separated by a factor of about $5 \times 10^{60}$: the universe, the Planck scale and what we call the sub Planck scale. Integration of the Einstein cosmological equations, for this fractal universe, gives the solution of a non-expanding universe with the present value of the observed numerical parameters. The red shift measured from the distant galaxies is interpreted here as due to the decreasing speed of light in a fractal universe.

## 1.-Introduction

Much work has been done on the fractality of space-time. See for example Mandelbrot (1982), Ord (1983), Nottale (1993), El Naschie (2004) etc. Here we start with the assumption of the fractality of time and apply its property of self-similarity. This means that any time interval of any clock is proportional to the age of the universe. In that case the ratio of these two times is a constant, a universal constant that defines the particular clock with its particular ticking. We analyse a few clocks and find out that space is fractal too. Adding the usual assumption that Planck's constant $\hbar$ is a true universal constant (as a consequence of the constancy of angular momentum), we arrive at the important conclusion that all lengths in the universe are constants, including the size of the universe. Then there is no expansion. Obviously the speed of light, as any other speed, decreases with time due to the ticks of the clocks being proportional to the age of the universe. We have presented elsewhere the proportionality between the speed of light and the Hubble parameter, Alfonso-Faus (2008).

As one would expect from relativity it turns out that space is also fractal. The constancy of Planck's constant $\hbar$ is also a plausible assumption as a result of the lifetime found for beta decay: this interval of time is proportional to the 7<sup>th</sup> power of ℏ so that a small variation in time for ℏ would certainly have to be noticed. Then we get the validity of the constancy of two well known physical properties (if no interaction is present): the angular momentum and the linear momentum. At a cosmological scale this corresponds to the cosmological principle of homogeneity and isotropy.

Zel'dovich (1967) formulation for the cosmological constant Λ and Weinberg (1972) relation between quantum and cosmological parameters have been shown to be one and the same thing, Alfonso-Faus (2008). They have also been shown to be a result of Newton's laws, Alfonso-Faus (2008). We note that these relations are valid for the cosmological scale, the quantum scale, as well as for the Planck and sub Planck scales.

Finally we integrate the Einstein cosmological equations for the case we are dealing with. We use the most current values for the Ω parameters, $\Omega_m = 1/3$, $\Omega_\Lambda = 2/3$ and $w = -1$ for the equation of state.

## 2.-Fractal time

Fractality implies the property of self-similarity. Applied to the dimension of time this means that any time interval $\delta t$ is proportional to the age of the universe t, $\delta t/t$ = dimensionless constant. This defines a scale and $\delta t$ defines the ticking of a clock, any type of clock. We establish the proportionality with t choosing the scales of Planck, atomic, inverse photon frequency and gravitational period (Kepler):

$$\left(\frac{G\hbar}{c^5}\right)^{1/2} \propto \frac{\hbar}{mc^2} \propto \frac{\lambda}{c} \propto \left(\frac{a^3}{Gm}\right)^{1/2} \propto t \qquad (1)$$

We now take the usual assumption that ℏ is constant. We also take the assumption that the linear momentum mc is constant. Then from (1) we obtain

$$\begin{aligned} ct &= const \\ \frac{G}{c^3} &= const \\ a &= const \end{aligned} \qquad (2)$$

We know that in order to derive the Einstein's field equations from the action principle, the factors in from of the integrals must be constant. These are $G/c^3$ and $mc$. The constancy of $mc$, the linear momentum, is a very well known law. The constancy of the mass rate $c^3/G$ is not so well known and its physical implications should not be overlooked, Alfonso-Faus (2008).

**3.-Fractal space**

Having analysed the consequences of the assumption of a fractal time, we now see that the fractality of space is inevitable too. All we have to do is to multiply the dimension time in (1) by the speed of light c, and of course note that from (2) the product ct is a constant. Then Planck's length is constant, so is the size of the atomic sizes, as well as the wavelengths of photons. Fractality of space implies a constant size for everything in the universe. And the immediate conclusion is that all speeds, including c, decrease inversely proportional to the cosmological time. And that all masses increase proportionally to t, the Mass-Boom effect, Alfonso-Faus (2008). In addition there is no expansion of the universe. The observed Hubble red shift is seen here as a result of the time variation of the speed of light.

**4.-Fractal mass**

The constancy of the linear momentum, together with the linear decrease of the speeds, implies the linear increase of all masses. This means that the ratio of two different masses is a universal constant. This is a self-similarity property of fractality. If M is the mass of the universe and t its age then the ratio t/M is a universal constant. Clearly it is the universal constant $G/c^3$ in (2), an expression of Mach's principle. We then have a fractal universe: fractal space-time <u>and fractal mass.</u>

**5.-Universal constants**

One expects, as Dirac noticed in 1937, that all universal constants at the cosmological level must be of order one. We can define the size of the universe, its total momentum content averaged over t, its total angular momentum and the mass rate as universal constants of order one:

$$\begin{aligned} ct &= 1 \\ Mc &= 1 \\ Mc.ct &= 1 = H_u \\ \frac{c^3}{G} &= 1 = \frac{M}{t} \end{aligned} \qquad (3)$$

All these properties refer to the scale of the universe. We recover the Planck's scale by applying to the mass, length and time the dividing constant factor $5\times10^{60}$. By dividing once more using the same factor we get the sub Planck scale with mass $10^{-65}$ gr., length $10^{-94}$ cm and tick $10^{-105}$ sec. Zel'dovich (1967) and Weinberg (1972) relations hold in these scales and the quantum one.

## 6.-Cosmological equations

The Einstein cosmological equations for a constant size universe and a total pressure $p = w\rho_m c^2$ plus $p_Q$ are given by

$$8\pi \frac{Gp}{c^2} + \frac{kc^2}{R^2} = \Lambda c^2$$
$$-\frac{8\pi}{3} G\rho_m + \frac{kc^2}{R^2} = \frac{\Lambda c^2}{3} \qquad (4)$$

With the usual $\Omega$ definitions these equations become

$$3w\Omega_m + 3\Omega_Q + \Omega_k = 3\Omega_\Lambda$$
$$-\Omega_m + \Omega_k = \Omega_\Lambda \qquad (5)$$

which have the solution $w = -1$, $\Omega_m = 1/3$, $\Omega_k = 1$, $\Omega_\Lambda = \Omega_Q = 2/3$. The parameter $\Omega_Q$ may be identified for example to an electrical outward pressure due to the fact that the Debye length for the universe is a bit larger than its size ct. An overall net charge Q, of the order of $5\times10^{60}$ $e$, may be present in the universe, $e$ the charge of the electron.

## 7.-Conclusions

Much work has been done on the fractality of space-time. Here we add the fractality of mass. The universal constants at the cosmological level are of order one and the physical properties of mass and time may be considered as identical. Then the universe is fractal. All speeds, including the speed of light, decrease with the cosmological time, and the length travelled in each tick of the clock is constant. Then the size of the universe as given by ct is constant. The horizon problem is not present in this theory, neither the entropy nor the cosmological constant problem, etc. The so called coincidences are just an effect of the scaling factors.

## 8.-References

Alfonso-Faus, A., (2008), Astrophysics and Space Science, 315, 25 and next issue. Also in the arXiv.org archives.

El Naschie, M.S., (2004), in "Chaos, Solitons and Fractals" Vol 19, is.1, January.

Mandelbrot, Benoit B, (1982), "The fractal geometry of nature", W.H. Freeman and Co.

Nottale L., (1993), "Fractal Space-Time and Microphysics", World Scientific Pub. Co. Inc., May.

Ord, G.N., (1983), Phys. A : Math. Gen16, 1869-1884.

Weinberg, S., (1972), "Gravitation and Cosmology", John Wiley and Sons, New York.

Zel'dovich, Ya.B., (1967), Pis'ma Zh. Eksp. Teor. Fiz., vol 6, p. 883.